\documentclass[oupdraft]{bio}
\usepackage{url}

\usepackage[english]{babel}
\usepackage[utf8x]{inputenc}
\usepackage[T1]{fontenc}
\usepackage{xcolor}
\usepackage{bbm}
\usepackage{mathtools}
\history{Received August 1, 2010;
revised October 1, 2010;
accepted for publication November 1, 2010}

\newcommand{\mS}{\mathcal{S}}
\newcommand{\mT}{\mathcal{T}}
\newcommand{\etab}{\boldsymbol{\eta}}
\newcommand{\bx}{{\bf x}}
\newcommand{\zetab}{\boldsymbol{\zeta}}
\newcommand{\betab}{\boldsymbol{\beta}}
\newcommand{\vect}{\text{vec}}

\begin{document}

\title{Bayesian Analysis of Multidimensional Functional Data}

\author{JOHN SHAMSHOIAN, DAMLA \c{S}ENT\"{U}RK\\[4pt]
\textit{Department of Biostatistics, University of California Los Angeles,
	 650 Charles E Young Drive, Los Angeles, CA, 90095, USA}
\\[2pt]
SHAFALI JESTE\\[4pt]
\textit{Department of Psychiatry and Biobehavioral Sciences,
University of California Los Angeles,757 Westwood Plaza, Los Angeles,
CA, 90095, USA}\\[2pt]
DONATELLO TELESCA$^\ast$\\[4pt]
\textit{Department of Biostatistics, University of California Los Angeles,
	 650 Charles E Young Drive, Los Angeles, CA, 90095, USA}\\[2pt]
{dtelesca@ucla.edu}}

\markboth%
{J. Shamshoian and others}
{Bayesian Analysis of Multidimensional Functional Data}

\maketitle

\footnotetext{To whom correspondence should be addressed.}

\newpage
\begin{abstract}
{Multi-dimensional functional data arises in numerous modern scientific experimental and observational studies. In this paper we focus on longitudinal functional data, a structured form of multidimensional functional data. Operating within a longitudinal functional framework  we aim to  capture low dimensional interpretable features. We propose a computationally efficient nonparametric Bayesian method to simultaneously smooth observed data, estimate conditional functional means and functional covariance surfaces. Statistical inference is based on Monte Carlo samples  from the posterior measure through adaptive blocked Gibbs sampling.  Several operative characteristics associated with the proposed modeling framework are assessed comparatively in a simulated environment.  We illustrate the application of our work in two case studies.  The first case study involves age-specific fertility collected over time for various countries. The second case study is an implicit learning experiment in children with Autism Spectrum Disorder (ASD). }
{Functional data analysis; Rank regularization; Tensor spline; Factor analysis; Longitudinal mixed model; Gaussian process; Marginal covariance}
\end{abstract}

\newpage
\setcounter{page}{1}
\section{Introduction}
\label{s:intro}

Many modern biomedical experiments result in \emph{functional data};  data that are realizations from a continuous stochastic process defined over functions of a specific evaluation domain. Typically, each statistical unit contributes one or several random functions, and a sample of $n$ such functions is collected for statistical analysis.   In this paper we investigate modeling and inference for longitudinal functional data, conceptualized as functional data observed repeatedly over several longitudinal time-points. A typical  dataset would contain $n$ patients observed over the course of multiple visit times,  with each visit contributing a functional datum. Thus, for patient $i$ we would record the outcome $y_{i}(s,t)$, where $s$ is the visit time and $t$ is the functional argument. In this setting it is reasonable to expect   non-trivial correlations between functions from one visit time to another.  Therefore, appropriate modeling of this dependence pattern would be critical for the validity of statistical inference. 
	 
When modeling structured functional data, important progress can be made by leveraging simplifying assumptions about the data sampling process, and  by characterizing  high-dimensional dependence  through lower dimensional structures.  This general approach has received considerable attention in the literature.  In the setting of repeated functional measurements, \cite{di2009} introduced the hierarchical functional ANOVA. In longitudinal settings, \cite{Greven2010} proposed a decomposition based on a functional random intercept and slope to capture longitudinal variations. This approach has been extended in \cite{Chen2012}, through the use of data-driven time-varying basis functions, employing functional principal components analysis (FPCA) at every longitudinal time point. By contrast \cite{Staicu2015},  and similarly \cite{Chen2015},  developed a theory and justification for using a more parsimonious data-driven time-invariant basis functions obtained through marginalization of covariance operators. The appealing nature and flexibility of structured FPCA modeling strategies has seen  the application and extension of these methods  to   challenging scientific problems ranging from functional brain imaging (\citealt{Hasenstab2017, Scheffler2018}),  to the exploration of complex data from wearable devices (\citealt{Goldsmith2016}).

The vast majority of approaches based on FPCA, generally focus on point estimation from a frequentist perspective, and do not provide reliable uncertainty quantification without bootstrapping. The very application of the bootstrap methodology to structured functional data  has not been the subject of rigorous investigation. The literature, in fact, is ambiguous on the handling of the many tuning parameters, typical of structured FPCA models.  Although there are some consistency results regarding the bootstrap for functional data (\citealt{Cuevas2006, Ferraty2010}), the procedure is relatively underdeveloped for hierarchical data (\citealt{Ren2010}).  

Bayesian methods in functional data analysis define a straightforward mechanism for uncertainty quantification. 
This appealing inferential structure comes, however, at the cost of having to specify a full probability model and priors with broad support on high dimensional spaces. When random functions are modeled as realizations of Gaussian processes,  modeling covariance operators for one-dimensional random functions often relies on  semi-parametric  assumptions (\citealt{Shi2011}). More flexible representation have been proposed by \cite{Cox2016} and \cite{Cox2017},  who use inverse Wishart   process priors. This process, arising as an extension of the finite dimensional inverse Wishart distribution, tend to inherit its limitations  and  often result in unwarranted prior bias on correlation components. In hierarchical and multi-dimensional functional data settings,  starting from the seminal work of \cite{Morris2009},  and recent extensions in \cite{Lee:2019aa}, the prevalent strategy has been to work within the framework of basis function transforms, defining flexible mixed effect models at the level of the basis coefficients. The resulting functional mixed effects models, like their finite dimensional counterpart, require a certain degree of subject matter expertise in the definition of random effects and their covariance structure. Furthermore, specific implications about the form of the marginal covariance are often non data-adaptive in highly structured settings.

This manuscript aims to merge the appealing characterization of longitudinal functional data through FPCA decompositions (\citealt{Chen2012, Staicu2015, Chen2015}), with flexible probabilistic representations of the classical Karhunen-Lo\'{e}ve expansion of square integrable random functions.  Our work builds on the ideas of \cite{Suarez2017} and \cite{Montagna2012}, who adapted the regularized product Gamma prior for infinite factor models of \cite{Dunson}, to the analysis of random functions. Extensions of this framework to the longitudinal functional setting are discussed in Section \ref{s:model}.   In Section \ref{s:prior} we discuss prior distributions and ensuing implications for the covariance operator.  A comprehensive framework for posterior inference is discussed in Section \ref{s:posterior}.   Section \ref{s:simulation} contains a comparative simulation study. Finally, in  Section \ref{s:data} we discuss the application of our proposed methodology to two case studies. The first case study explores age-specific fertility dynamics in the global demographic study conducted by the Max Plank Institute and the Vienna Institute of Demography (\citealt{HFD}). While purely illustrative, this data allows for a direct comparison with the original analysis of  \cite{Chen2015}. The second case study, involves the analysis of electroencephalogram (EEG) data from an investigation of implicit learning in children with autism spectrum disorder (ASD) (\citealt{Jeste}).

\section{A Probability Model for Longitudinal Functional Data}
\label{s:model}
	
Let $y_{i}(s, t)$ denote the response for subject $i$, ($i = 1,\ldots, n$), at longitudinal time $s\in\mS$ and functional time $t\in \mT$, where  $\mS$ and $\mT$ are compact subspaces of $\mathbb{R}$. In practice, we only obtain observations  $y_i(s_j, t_k)$ at discrete sampling locations $(s_{j}, t_{k})\in \mS\times\mT$, $j = 1,\ldots,n^{s}_{i}$, $k = 1,\ldots,n^{t}_{i}$. However, in subsequent developments, we maintain the lighter notation $y_i(s,t)$ without loss of generality.  

Let $f_i(s,t)$ be a Gaussian Process (GP) with mean  $E\{f_i(s,t)\} = \mu(\bx_i, s, t)$, possibly dependent on  covariate information $\bx_i\in \mathbb{R}^d$, and covariance kernel $Cov\{y_i(s,t), y_i(s',t')\} =$\\ $K\{(s,t),(s',t')\}$.  	
A familiar sampling model for $y_i(s,t)$ assumes:
\begin{equation}
 y_{i}(s, t) = f_{i}(s, t) + \epsilon_{i}(s,t),\;\; \epsilon_{i}(s, t) \stackrel{iid}{\sim} N(0, \varphi^{2});
\label{eq: sampling model}
\end{equation}
where $\varphi^2>0$ is the overall residual variance.  Given a set of suitable basis functions  $b^{(1)}_m(s): \mS \rightarrow \mathbb{R}$, $(m=1,2,\ldots p_1)$, and $b^{(2)}_\ell(t):\mT\rightarrow\mathbb{R}$, $(\ell=1,2,\ldots p_2)$, and a set of random coefficients $\theta_{im\ell}$, the prior for the underlying signal $f_{i}(s,t)$ is constructed through a random tensor product expansion, so that 
\begin{align*}
     f_{i}(s,t) &= \sum_{m = 1}^{p_{1}}\sum_{l=1}^{p_{2}}b_{m}^{(1)}(s)b^{(2)}_{l}(t)\theta_{iml}.
\end{align*}	
Since the truncation values $p_{1}$ and $p_{2}$ may be large to insure small bias in the estimation of the true $f_i(s,t)$, we  follow \cite{Dunson} and project the basis coefficients on a lower dimensional space.

Let $\Theta_{i} = \{\theta_{im\ell}\}\in \mathbb{R}^{p_{1}\times p_{2}}$ be the matrix of basis coefficients for subject $i$. 
After defining loading matrices $\Lambda\in\mathbb{R}^{p_1\times q_1}$, $(q_1 \ll p_1)$, and $\Gamma\in\mathbb{R}^{p_2\times q_2}$, ($q_2 \ll p_2$), and a latent matrix of random scores $\etab_i\in \mathbb{R}^{q_1\times q_2}$, we assume
 \begin{equation}
    \Theta_{i} = \Lambda\etab_{i}\Gamma^{\top} + \zetab_{i},\;\; \vect(\zetab_{i}) \sim \mathcal{N}(0, \Sigma);
\label{eq: factor model}	    
\end{equation}
where $\Sigma$ is taken to be diagonal.  The foregoing construction has connections with factor analysis. In fact, vectorizing $\Theta_i$ we obtain
\begin{align*}
    \vect(\Theta_{i}) = (\Gamma\otimes\Lambda)\vect(\etab_{i}) + \vect(\zetab_{i});
\end{align*}
which resembles the familiar ($q_1\times q_2$) latent factor model, with  loading matrix $\Gamma\otimes\Lambda$
and latent factors $\vect(\etab_i)$. Differently from standard latent factor models, our use of a  Kronecker product representation for the loading matrix introduces additional structural assumptions about $Cov(\Theta_i)$,  and the ensuing form of the covariance kernel  $K\{(s,t),(s',t')\}$. 

More precisely, assuming $Cov(\etab_i) = H$, the marginal covariance of $\Theta_i$ takes the form
\begin{equation}
Cov(\Theta_i) = (\Gamma \otimes \Lambda)H(\Gamma\otimes \Lambda)^{\top} + \Sigma  = \Omega.
\label{eq: covariance theta}
\end{equation}
Furthermore, defining $B_{1}(s) = \left(b^{(1)}_{1}(s),\ldots, b^{(1)}_{p_{1}}(s)\right)^{\top}$ and $B_{2}(t) = \left(b^{(2)}_{1}(t),\ldots,b^{(2)}_{p_{2}}(t)\right)^{\top}$,  induces the following representation for the covariance kernel $K\{(s,t),(s',t')\}$,
\begin{equation}
\label{eq: covariance kernel}
K\{(s,t),(s',t')\} = \{B_{1}(s)\otimes B_{2}(t)\}\,\Omega\, \{B_{1}(s')\otimes B_{2}(t')\}^{\top}.
\end{equation}
The low-rank structure of $\Omega$ in ({\ref{eq: covariance theta}}), depends on the number of latent factors $q_1$ and $q_2$ in the quadratic form $(\Gamma \otimes \Lambda)^{\top}\,H\,(\Gamma\otimes \Lambda)$. Rather than selecting the number of factors a priori, 
in Section \ref{s:prior} we introduce prior distributions encoding rank restrictions through continuous stochastic regularization of the loading coefficient's magnitude.
Additional structural restrictions may ensue from specific assumptions about the latent factors covariance $H$. Specifically, setting $H = I_{q_{1}q_{2}}$ leads to strong covariance separability of the longitudinal and functional dimensions. A more flexible covariance model hinges on the notion of weak-separability (\citealt{Chen2017}). This is achieved by setting $H=\mbox{diag}(h_1,\ldots, h_{q_1q_2})>0$.

Finally, let $\bx_{i}$ be a $d$-dimensional time-stable covariate for subject $i$. Dependence of the longitudinal functional outcome $y_i(s,t)$ on this set of predictors is conveniently introduced through the prior expectation of $\etab_i$. More precisely, let $\betab$  be a $d\times q_{1}q_{2}$ matrix of regression coefficients, we assume
\begin{align*}
    \vect(\etab_{i})  \sim N(\betab^{\top}\bx_{i},\, H),
\end{align*}
which implies the following marginal mean structure for $y_i(s,t)$,
\begin{equation}
\label{eq: mean structure}
 E\{y_i(s,t)\} =  \mu(\bx_{i},s,t) = \{B_{1}(s)\Gamma \otimes B_{2}(t)\Lambda\}\,\betab^{\top}\bx_{i}.
\end{equation}
The model in (\ref{eq: sampling model}), together with the sandwich factor construction in (\ref{eq: factor model}) defines a probabilistic representation of the product FPCA decomposition in \cite{Chen2015}. An intuitive parallel is introduced in Section \ref{s:prior}, and a technical discussion is provided in the accompanying web-based supplementary document. Differently from \cite{Chen2015}, we propose model-based inference through regularized estimation based  on the posterior measure.

\section{Rank Regularization and Prior Distributions}
\label{s:prior}

The selection of prior distributions for all parameters introduced in Section \ref{s:model} is guided by the following considerations. Let $\gamma_{\ell j}$ and $\lambda_{mk}$ be specific entries in the loading matrices $\Gamma$ and $\Lambda$ respectively. Defining $\psi_{j}(s) = \sum_{l = 1}^{p_{1}}\gamma_{lj}b^{(1)}_{l}(s)$ and $\phi_{k}(t) = \sum_{m=1}^{p_{2}}\lambda_{mk}b^{(2)}_{m}(t)$, we may expand  $f_{i}(s,t)$ as follows:
\begin{align*}
    f_{i}(s,t) &= \sum_{j = 1}^{q_{1}}\sum_{k = 1}^{q_{2}}\psi_{j}(s)\phi_{k}(t)\eta_{ijk} + r_{i}(s,t),\\
    r_{i}(s,t) &= \sum_{j=1}^{p_{1}}\sum_{k=1}^{p_{2}}b^{(1)}_{j}(s)b^{(2)}_{k}(t)\zeta_{ijk}.
\end{align*}
The first  component in the expression for $f_i(s,t)$ describes a mechanism of random functional variability which depend on the tensor combination of $q_1$ and $q_2$ data-adaptive basis functions $\psi_j(s)$ and $\phi_k(t)$ respectively, and $q_1 \times q_2$ basis coefficients $\eta_{ijk}$. Given $q_1$ and $q_2$, any residual variability is represented in the random function $r_{i}(s,t)$. When $\psi_j(s)$ and $\phi_k(t)$ are chosen to be eigenfunctions of the marginal covariance kernels in $s$ and $t$, this representation is essentially equivalent to the  product FPCA construction of \cite{Chen2015}. 
	
Statistical inference for FPCA constructions, commonly selects a small number of eigenfunctions on the basis of empirical considerations. Here we take an adaptive regularization approach, choose $q_1$ and $q_2$ relatively large, and   assume the variance components in $\Lambda$ and $\Gamma$ to follow a modified multiplicative gamma process prior (MGPP) \cite{Dunson}  \cite{Montagna2012}.  

Let $\lambda_{mk}$   denote the $(m,k)$ entry of $\Lambda$. The modified MGPP is defined by setting  
\begin{align*}
    \lambda_{mk} &\sim N(0, \rho_{1mk}^{-1}\tau_{1k}^{-1}),\hspace{.3cm} \rho_{1mk}^{-1} \sim \text{Ga}(\nu_{1}/2,\nu_{1}/2),\\
	\tau_{1k} &= \prod_{\upsilon=1}^{k}\delta_{1\upsilon}, \hspace{.3cm}\delta_{11} \sim \text{Ga}(a_{11}, 1),\hspace{.3cm} \delta_{1\upsilon} \sim \text{Ga}(a_{12}, 1)\mathbbm{1}(\delta_{1\upsilon} > 1), \mbox{ for } \upsilon\geq 2; \;\; k= 1,2,\ldots,q_1.
\end{align*}
This prior is designed to encourage small loadings in $\Lambda$ as the column index increases. In the original formulation of \cite{Dunson} and \cite{Montagna2012}, choosing $a_{12} > 1$, insures stochastic ordering of the prior precision, in the sense that $E(\tau_{1k}) < E(\tau_{1(k+1)})$, for any $k=1,2,\ldots,(q_1-1)$. In our setting, we require the more stringent probabilistic ordering  $Pr(\tau_{1k} < \tau_{1(k+1)}) = 1$, by assuming $\delta_{1\upsilon} > 1$, which results in a more stable and efficient Gibbs sampling scheme. Analogous regularization over the columns of $\Gamma$ is achieved by setting: 
\begin{align*}
    \gamma_{lj} &\sim N(0, \rho_{2lj}^{-1}\tau_{2l}^{-1}), \hspace{.3cm}\rho_{2lj}^{-1}\sim\text{Ga}(\nu_{2}/2,\nu_{2}/2)\\
    \tau_{2l} &= \prod_{\upsilon=1}^{l}\delta_{2\upsilon}, \hspace{.3cm}\delta_{21} \sim \text{Ga}(a_{21}, 1),\hspace{.3cm} \delta_{2\upsilon} \sim \text{Ga}(a_{22}, 1)\mathbbm{1}(\delta_{2\upsilon} > 1), \mbox{ for } \upsilon \geq 2; \;\; l=1,2,\ldots,q_2.
\end{align*}
Adaptive shrinkage is induced by placing hyper-priors on $a_{11}, a_{12}, a_{21},$ and $a_{22}$, such that
	$$
	    a_{11}, a_{21} \stackrel{ind}{\sim} \text{Gamma}(r_{1}, 1), \;\;
	    a_{12}, a_{22} \stackrel{ind}{\sim} \text{Gamma}(r_{2}, 1).
	$$
The model is completed with priors on residual variance components and regression coefficients.
Specifically, conditionally conjugate priors are placed on the diagonal elements of $\Sigma$ and $H$, respectively, as well as the residual variance $\varphi$, such that: 
$$
 \sigma_{j}^{-1} \sim \text{Ga}(a_{\sigma},b_{\sigma}), \;\;
	    h_{j}^{-1} \sim \text{Ga}(a_{h}, b_{h}),\;\;
	    \varphi^{-1} \sim \text{Ga}(a_{\varphi}, b_{\varphi}).
$$
Finally, we induce a Cauchy prior for the regression coefficients matrix $\betab$ as in \cite{Montagna2012}. Denoting with $\beta_{j\ell}$ the $(j,\ell)$ entry of $\betab$, we assume
$$
 \beta_{j\ell} \sim N(0, \omega_{j\ell}), \;\; \omega_{j\ell}^{-1} \sim \text{Ga}(1/2, 1/2); \hspace{.3cm} \ell = 1,\ldots,q_{1}q_{2},\hspace{.3cm}j = 1,\ldots,d.\\
$$

\section{Posterior Inference}
\label{s:posterior}
	
Posterior simulation through Markov chain Monte Carlo is relatively straightforward, after selection of an appropriate basis transform and truncation of $\Gamma$ and $\Lambda$ to include $q_1\ll p_1$  and $q_2\ll p_2$ columns respectively.  The use of conditionally conjugate priors  allows for simple Gibbs transitions for all parameters, with the exceptions of the shrinkage parameters $a_{11}, a_{12}, a_{21}, \text{ and } a_{22}$, which are updated via a Metropolis-Hastings step. A detailed description of the proposed algorithm is reported in the web-based supplement. 

We note that the decomposition of $Cov(\Theta_i)$ in (\ref{eq: covariance theta}) may not be unique. However, from a Bayesian perspective, one does not require identifiability of the loading elements for the purpose of covariance  estimation.  Direct inference for  $K\{(s,t),(s',t')\}$ and its functionals may be achieved by post-processing Monte Carlo draws from the posterior $p(\Omega\mid {\bf y})$ and evaluating the covariance function over arbitrarily dense points ${\bf t}^* \coloneqq (t_1^*,\ldots, t_{w1}^*)^\top \in \mT$  and ${\bf s}^* \coloneqq (s_1^*,\ldots,s^*_{w2})^\top\in \mS$ using (\ref{eq: covariance kernel}). Analogously, given samples from $p(\betab\mid {\bf y})$, inference about the mean structure is achieved evaluating  $\mu(\bx_i, s, t)$ over ${\bf s}^*$ and  ${\bf t}^*$ using the expansion in (\ref{eq: mean structure}).  

Even though Monte Carlo samples are easily obtained, exploring  the posterior measure for a four-dimensional object like $K\{(s,t)(s',t')\}$ may still be be daunting. Some useful posterior summaries may be obtained through marginalization. In particular, we may define marginal covariance functions $K_T(t,t')$ and $K_S(s,s')$ as follows:
\begin{equation}
\label{eq: marginal covariance}
\begin{array}{lclclcl}
K_T(t,t')  &=& \displaystyle\int_\mS K\{(s,t)(s,t')\} ds, &&
K_S(s,s') &=& \displaystyle\int_\mT K\{(s,t)(s',t)\} dt.
\end{array}
\end{equation}
Intuitively, $K_S(\cdot)$ and $K_T(\cdot)$ summarize patterns of functional co-variation along a specific coordinate, and their lower-dimensional posterior summaries may be obtained through functional eigenanalysis as in \cite{Chen2015}. For example,  posterior draws from $K_{S}(s,s')$,  can be approximated numerically by averaging over a fine evaluation grid ${\bf t}^*$, s.t.
\begin{align*}
    K_{S}(s,s') &\approx \frac{1}{w_1}\sum_{\ell=1}^{w_1}K\{(s,t^*_\ell),(s',t^*_\ell)\}.
\end{align*}
For each Monte Carlo sample, a spectral analysis of $K_S(s,s')$ yields eigenfunctions $\tilde{\psi}_j(s)$, $(j=1,2,\infty)$,  summarizing the posterior functional principal components. Because eigenfunctions are determined up to $\pm$ sign,  some care must be taken when defining posterior summaries.  To handle the sign ambiguity in the definition of ergodic averages, we store a running mean of each eigenfunction over Monte Carlo iterations.  As we move through the sample, we multiply by $-1$ if the sign-switched eigenfunction has smaller squared distance from the running mean. Posterior summaries are therefore defined over appropriately oriented components.  Finally, simultaneous credible intervals for all functions of interest are easily obtained from Monte Carlo samples, by applying the methodology discussed in \cite{Crainiceanu}. 

\vskip.1in
\noindent The proposed modeling framework relies on a specific basis transform strategy. While the literature has suggested the use of zero-loss transforms as a default option (\citealt{Morris2009, Lee:2019aa}), we find that it is not uncommon to observe some sensitivity to the number of basis functions used in the initial projection. Furthermore,  the choice of more parsimonious designs, when warranted by the application, may lead to important gains in computational and estimation efficiency.  	

We propose to select the number of basis functions as a fraction of the original sampling intensity. An optimized search amongst a set of candidate bases may simply rely on the minimization of information criteria. We consider simple versions of  the deviance information criterion (DIC), and  Bayesian information criteria 1 \& 2 (BIC 1 \& BIC 2), counting only fixed or both fixed and random effects (\citealt{delattre2014}).   A detailed description  of these summaries and their computation is provided in the web-based supplement.

\section{A Monte Carlo Study of Operating Characteristics}
\label{s:simulation}
We performed a series of numerical experiments aimed at evaluating the estimation performance for both the functional mean and covariance.  We study three simulation scenarios, including two weakly separable kernels  (cases 1 and 2) and one non-separable covariance function (case 3). Specifically,  for $s\in [0,1]$ and $t\in [0,1]$, we take:
\begin{enumerate}
\item $K_{\mathcal{S}}(s,s') = \sum_{j=1}^{2}\lambda_{j}\psi_{j}(s)\psi_{j}(s')$, with eigenvalues $\lambda_{j} = \frac{1}{j^{2}\pi^{2}}$ and eigenfunctions \linebreak $\psi_{j}(s) = \sqrt{2}\sin(j\pi s)$, $K_{\mathcal{T}}(t,t') = \sigma^{2}\bigg(1 + \frac{\sqrt{3}|t-t'|}{\rho}\bigg)\exp\bigg(-\frac{\sqrt{3}|t-t'|}{\rho}\bigg)$, in the Mat\`{e}rn class, and  mean $\mu(s,t) = \sqrt{\frac{1}{5\sqrt{s+1}}}\sin(5t)$. 
\item $K_{\mathcal{S}}(s,s') = \sum_{j=1}^{2}\lambda_{j}\psi_{j}(s)\psi_{j}(s')$, with eigenvalues $\lambda_{j} = \frac{1}{(j-1/2)^{2}\pi^{2}}$ and eigenfunctions \linebreak $\psi_{j}(s) = \sqrt{2}\sin((j-1/2)\pi s)$,  $K_{\mathcal{T}}(t,t') = \sum_{k=1}^{50}\lambda_{k}\phi_{k}(t)\phi_{k}(t')$, with $\lambda_{k} = k^{-2\alpha}$ and $\phi_{k}(t) = \cos(k\pi t)$, and mean $\mu(s,t) = 5\sqrt{1-(s-.5)^{2} -(t-.5)^{2}}$. 
\item $\displaystyle K((s,t),(s',t')) = \frac{1}{(t-t')^{2}+1}\exp\bigg(-\frac{(s-s')^{2}}{(t-t')^{2}+1}\bigg)$, stationary non-separable (\citealt{Gneiting2002}), and mean $\mu(s,t) =\sqrt{1 + \sin(\pi s) + \cos(\pi t)}$. 
\end{enumerate}
After evaluating the marginal kernels  on 10 longitudinal time points and 20 functional time points, the simulation truth is obtained by projecting the analytical eigenfunctions onto a bivariate spline.  A detailed description of the  data generating process is reported in a web-based supplement. In all three scenarios we assume $y_{i}(s,t) \sim N[\mu(s,t), K\{(s,t),(s',t')\} + \varphi^{2}], \;(i = 1,\ldots n)$, with residual error variance set as $\varphi^{2} = .025$, and $n=30,60$. For fitting purposes we consider a model which is overparametrized relative to the truth and choose $B^{1}(t)$ to be cubic b-splines with knots at $t = (1/6, 2/6, 3/6, 4/6, 5/6, 5/6)$. $B^{2}(s)$ is also chosen to be  cubic b-splines with knots at $s = (1/5, 2/5, 3/5, 4/5)$. We also set $q_1 = \mbox{rank}(\Lambda)=6$ and  $q_2=\mbox{rank}(\Gamma)= 6$.  Finally, prior hyper-paramaters are set as follows: $\nu_{1} = 5$, $\nu_{2} = 5$, $r_{1} = 1$, $r_{2} = 2$, $a_{\sigma} = .5$, $b_{\sigma} = .5$, $a_{h} = 1$, $b_{h} = 1$, $a_{\varphi} = .0001$, and $b_{\varphi} = .0001$. 

We consider estimation of the mean, covariance, marginal covariance functions, and the associated two principal eigenfunctions. Each simulation includes 1,000 Monte Carlo experiments.   For each experiment, posterior estimates are based on 10,000 iterations of 4 independent Markov chains, after discarding 2,500 draws for burn-in. We compare estimation of covariance, marginal covariance functions, and associated two principal eigenfunctions to the respective estimates provided by the product FPCA (\citealt{Chen2015}), as well as finite-dimensional empirical estimates of the mean and covariance defined as by their vectorized sample counterparts. Estimates obtained  with the  product FPCA have data-type set to sparse and fraction of variance explained (FVE) threshold set to .9999. 

All comparisons are based on the relative mean integrated squared error. For a function $f$ with domain $D$ and estimator $\hat{f}$, we define $RE(\hat{f},f) = \int_{D}\{\hat{f}(u)-f(u)\}^{2}du / \int_{D}f(u)^2du$. Note that $D$ can be multi-dimensional and in practice the integral is replaced with a sum. 

Table \ref{t:MeanCovSims} compares mean $\mu(s,t)$ and covariance $K\{(s,t),(s',t')\}$ estimation under the three settings listed above.  We find that estimates from each method improve in accuracy with increasing sample size, with the posterior and product FPCA showing greater efficiency than empirical in terms of covariance estimation. Similar findings characterize the estimation performance of all marginal covariance functions ($K_{S}$, $K_{T}$), and the associated two principal eigenfunctions ($\psi_{i}(s)$, $i=1,2$), and ($\phi_{i}(t)$, $i=1,2$).  Detailed numerical results are reported in the web-based supplement. 
	
A small simulation aimed at assessing the performance of the information criteria proposed in Section \ref{s:posterior} is illustrated in  Table \ref{t:case 2 info}.  We considered the following  data-generating mechanism:  covariance case 2, 20 longitudinal points, 20 functional points, $N = 30$, $(p_{1}, p_{2}) = (10, 10)$, $(q_{1}, q_{2}) = (4, 4)$, and $\varphi^{2}=.025$. We fit candidate models with $(p_{1}, p_{2}) = (5,5), (10, 10),$ and $(15,15)$. We keep the number of latent factors as $(4, 4)$ in estimation, as the model is robust to the number of latent factors, due to adaptive penalization.  Table \ref{t:case 2 info} displays averaged information criteria over 1,000 simulations. The $(p_{1}, p_{2}) = (10, 10)$ row contains the smallest information criteria across all three metrics, giving strong indication that several alternative criteria tend to select an appropriate number of basis functions.

In summary, we observe that posterior estimates are associated with similar, and potentially improved efficiency in the estimation of the mean and covariance functions, when compared with product FPCA.  This similarity in estimation performance,  provides some empirical assurances that the chosen probabilistic representation of structured covariance functions, and estimation based on adaptive shrinkage, maintains a data-adaptive behavior with good operating characteristics. 

\section{Case Studies}
\label{s:data}
We illustrate the application of the proposed modeling frameworks in two case studies. The first dataset concerns fertility rate and age of mothers by country. The second case study focuses on functional brain imaging through EEG in the context of implicit learning in children with ASD. 

\subsection{Fertility rates}
\label{s: fertility}

The Human Fertility Database (\citealt{HFD}) compiles vital statistics to facilitate research on fertility in the past twentieth century and in the modern era. Age-specific fertility rates are available for 32 countries over different time periods. The age-specific fertility rate $ASFR(s,t)$ is defined as 
$$ASFR(s,t) = \frac{\text{births during year $s$ given by women aged $t$}}{\text{person-years lived during year $s$ by women aged $t$}}.$$
The dataset was previously analyzed and interpreted in a longitudinal functional framework using the product FPCA (\citealt{Chen2015}). This section focuses on a comparative analysis of product FPCA and the proposed probability model. 
	
\begin{figure}[!p]
	 	\centering
	 	\includegraphics[width=1\textwidth]{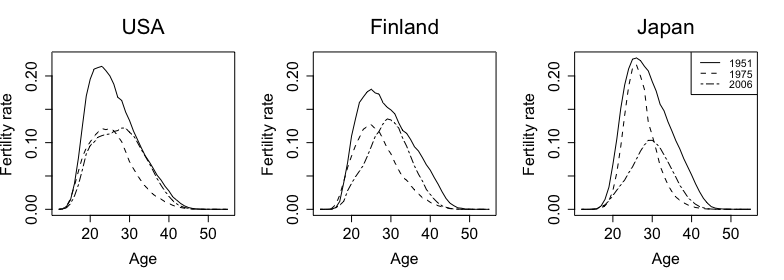}
	 	\caption{Raw age-specific fertility rate data for USA, Finland, and Japan. The displayed curves plot age-specific fertility data for years 1951, 1975, and 2006.}
	 	\label{fig:Data}
\end{figure}

We consider $n = 17$ countries, for the time period 1951 to 2006, 44 functional time points (ages 12-55), and 56 longitudinal time points (years 1951 to 2006). Figure \ref{fig:Data} illustrates three longitudinal functions for the USA, Finland, and Japan for years 1951, 1975, and 2006. The figure shows that overall fertility declined since 1951,  as well as a pattern of phase shifts in peak fertility towards later ages in later years. Since these rates are population measurements, we expect the data to contain very little noise. 

We use cubic b-splines as our basis functions since the data look smooth with no sharp changes in fertility rate over year or age of mother, and  consider $(p_{1}, p_{2}) = (22, 28)$ splines and $(q_{1}, q_{2}) = (11,10)$ latent factors, selected by minimizing the DIC.

\begin{figure}[!p]
	 	\centering
	 	\includegraphics[width=1\textwidth]{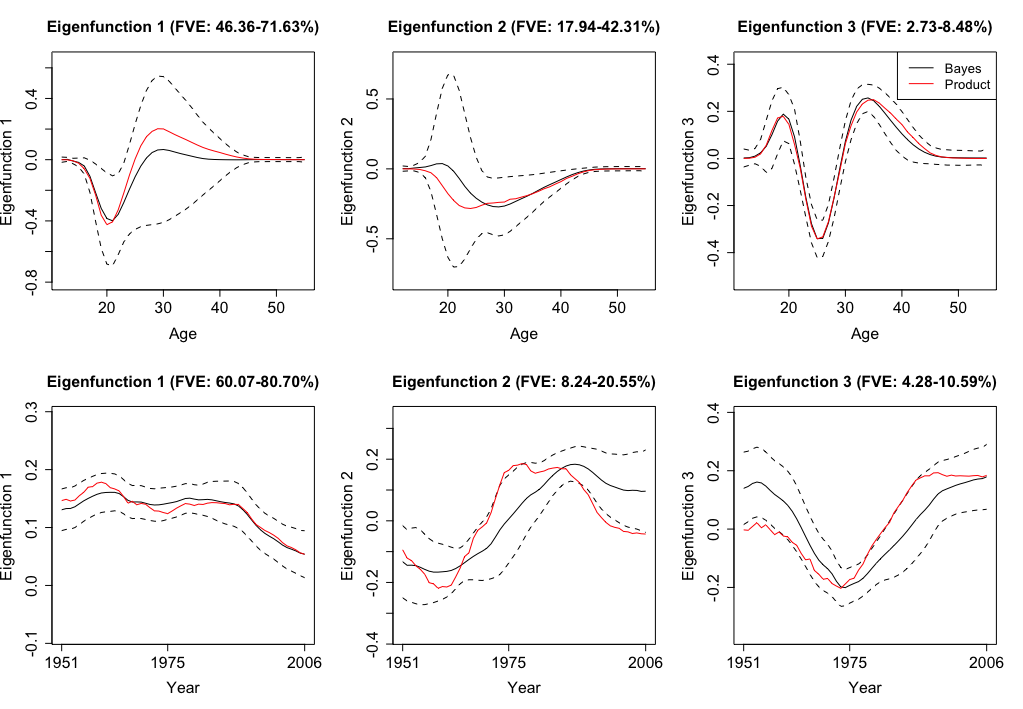}
	 	\caption{Age and calendar year marginal eigenfunctions. The above plots include the Bayesian posterior means, 95\% credible bands, and the product FPCA marginal eigenfunctions.}
	 	\label{fig:fertility_eigenfunctions}
	 \end{figure}

Longitudinal and aging dynamics are largely determined by their associated marginal covariance functions $K_{S}(s,s')$ and $K_{T}(t,t')$. Figure \ref{fig:fertility_eigenfunctions} displays the first three marginal eigenfunctions for age and calendar year. We include the  95\% simultaneous credible bands (\citealt{Crainiceanu}) as well as estimates obtained via product FPCA. We note that Bayesian posterior mean eigenfunctions are qualitatively similar to the inferred product FPCA estimates, therefore warranting similar interpretations to the one originally offered by \citealt{Chen2015}.  

In particular, the first marginal eigenfunction for age (Figure \ref{fig:fertility_eigenfunctions}, left panel) can be interpreted as the indexing variability in young fertility before the age of 25, with the second marginal eigenfunction for age (Figure \ref{fig:fertility_eigenfunctions}, central panel) indexing variability in fertility for mature age, between the ages of 20 and 40. As our modeling framework allows for rigorous uncertainty quantification in these posterior summaries, we note that the credible bands for the first and second eigenfunction are relatively wide, indicating that specific patterns should be interpreted with care.  
Examining directions of variance in fertility through the years, we note that the first marginal eigenfunction for year (Figure \ref{fig:fertility_eigenfunctions}, left panel) is relatively constant and can be interpreted as representing an overall ``size-component" of fertility from 1951-2006. The second eigenfunction  (Figure \ref{fig:fertility_eigenfunctions}, central panel) defines a contrasts of fertility in countries before and after 1975. For both the year and age coordinates the third marginal eigenfunctions capture a smaller fraction of the total variance and index higher patterns of dispersion at and around age 25 and at or around the year 1975.

\begin{figure}[!p]
    \hspace{-2cm}\includegraphics[width=1.2\textwidth]{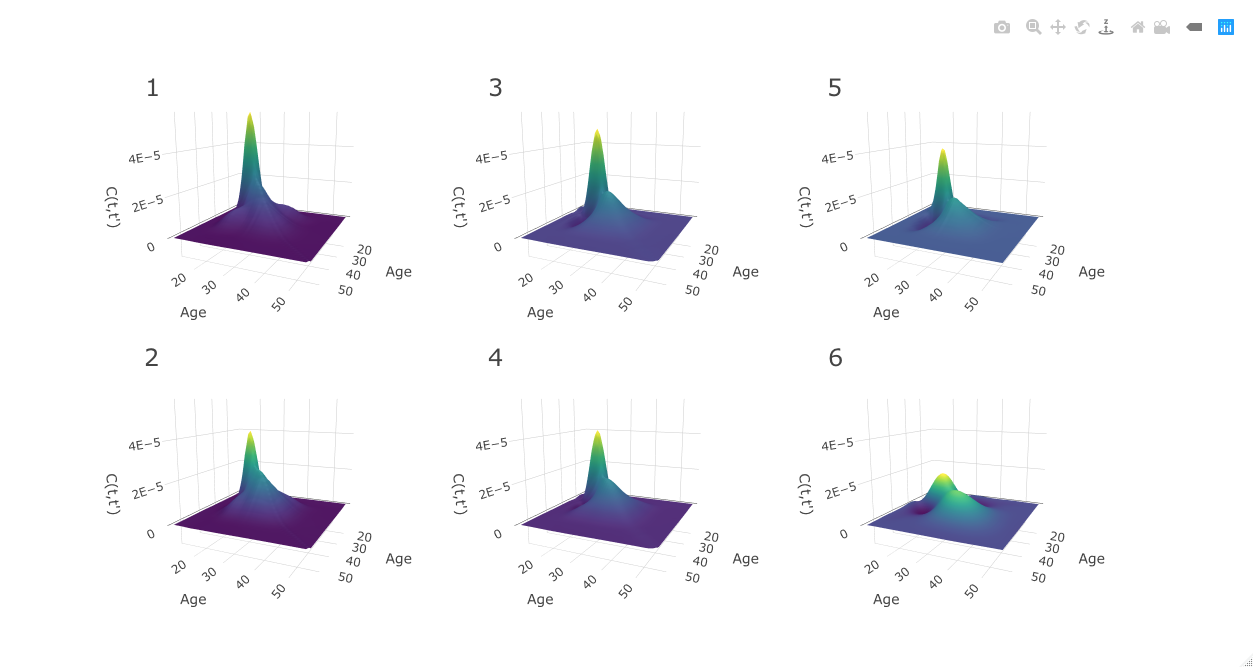}
    \caption{Sensitivity analysis for the marginal covariance function $K_{T}(t,t')$ (HFD study). Panels (1,2,3,4) refer to posterior mean estimates obtained under different projections     and numbers of latent factors (Specific details are provided in Section \ref{s: fertility}). Panels (5, 6) refer to product FPCA estimates obtained under dense (5) or sparse (6) settings.}
    \label{fig:marginal_func}
\end{figure}

We investigate sensitivity to the number of basis and latent factors considering four different models: model 1: $(p_{1}, p_2) = (44, 50)$, $(q_{1}, q_2) = (20, 20)$; model 2 $(p_{1}, p_2) = (44, 50)$, $(q_{1}, q_2) = (6, 6)$; model 3:  $(p_{1}, p_2)= (16,  20)$, $(q_{1},q_2) = (12, 12)$; and model 4: $(p_{1}, p_2) = (16, 20)$, $(q_{1}, q_2) = (6, 6)$. We also estimate the marginal covariance function with product FPCA using both the dense and sparse settings.  Point estimate for $K_T(t,t')$ are reported in Figure \ref{fig:marginal_func}.  Comparing estimates within column (left and center panels), we assess sensitivity to a drastic reduction in the  number of latent factors. Comparing estimates within row (left and center panels), we instead assess sensitivity to a drastic reduction in the number of basis functions. We note that the marginal age covariance function is relatively stable in all four settings. We contrast this relative robustness with  estimates based on the product FPCA. In particular, sparse estimation using 10-fold cross-validation results in meaningfully diminished local features. A possible reason for the instability is due to the small sample size (n=17).  In this example, Bayesian estimation is perhaps preferable, as adaptive penalization allows for stable estimates within a broad class of model specifications.

\subsection{An EEG Study on Implicit Learning in Children with ASD}
\label{s:ASD}
This analysis is motivated by a functional brain imaging study of implicit learning in young children with autism spectrum disorder (ASD), a developmental condition that affects an individual's communication and social interactions (\citealt{lord2000}). Implicit learning is defined as learning without the intention to learn or without the conscious awareness of the knowledge that has been acquired. 
We consider functional brain imaging through EEG, an important and highly prevalent imaging paradigm aimed at studying macroscopic neural oscillations projected onto the scalp in the form of electrophysiological signals. 

This study, carried out by our collaborators in the Jeste laboratory at UCLA, targets the neural correlates of implicit learning in the setting of an event-related shape learning paradigm (\citealt{Jeste}).  Children aged 2-6 years old with ASD were recruited through the UCLA Early Childhood Partial Hospitalization Program (ECPHP). Each participant had an official diagnosis of ASD prior to enrollment. Age-matched typically developing (TD) children from the greater Los Angeles area were recruited as controls. 

Six colored shapes (turquoise diamond, blue cross, yellow circle, pink square, green triangle, and red octagon) were presented one at a time in a continuous ``stream'' in the center of a computer monitor.  There were three shape pairings randomized to each child. For instance, a pink square may always be followed by a blue cross. After the blue cross would come a new shape pair. Within a shape pair would constitute an ``expected'' transition and between shape pairs would constitute an ``unexpected'' transition. Each child would wear a 128-electrode Geodesic Sensor Net and observe the stream of shapes on the computer monitor. Each stimulus, or presentation of a single shape, is referred to as a trial, and can result in frequency-specific changes to ongoing EEG oscillations, which are measured as Event Related Potentials (ERPs).
		  
Each waveform contains a phasic component called the P300 peak which represents attention to salient information. This phasic component 
is typically studied in EEG experiments and is thought to be related to cognitive processes and early category recognition (\citealt{Jeste}). We use the same post-processed data as in \cite{Hasenstab2017}. Namely, we consider 37 ASD patients and 34 TD patients using data from trials 5 through 60 and averaging ERPs in a 30 trial sliding window (\citealt{Hasenstab2015}). The sliding window enhances the signal to noise ratio at which the P300 peak locations can be identified for each waveform. Each waveform is sampled at 250 Hz resulting in 250 within-trial time points over 1000ms. Following \cite{Hasenstab2017}, we reduce each waveform to a 140ms window around each P300 peak. This 140ms window results in 37 within-trial time points. We do not apply warping techniques because each within-trial curve is centered about the P300 peak. Our analysis focuses on condition differentiation, formally defined as the difference between the expected and subsequent unexpected condition. Modeling condition differentiation for waveforms within a narrow window about the P300 peak over trials may give insights into learning rates for the ASD and TD groups. Thus, the main interest in this study is changes in condition differentiation over trials, and a longitudinal functional framework is required for statistical inference in this setting. Our analysis is based on the average condition differentiation within subject over the four electrodes in the right frontal region of the brain. 

We model the ASD and TD data cohorts separately, in order to estimate ERP time and trial covariance functions within group.  All inference is based on a model with $p_{1} = 20$, $p_{2} = 56$, $q_{1} = 10$, $q_{2} = 28$, selected minimizing DIC. A comprehensive analysis is reported in the web-based supplement. The number of MCMC iterations, burn-in, and hyper-parameters are set as in Section \ref{s: fertility}.

\begin{figure}[!p]
        \centering
        \includegraphics[width=.8\textwidth]{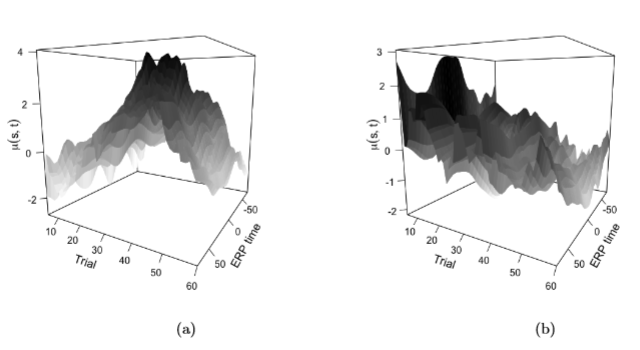}
        \caption{Posterior expected mean condition differentiation along trial and ERP time for the ASD (a) and the TD (b) cohorts.}
        \label{fig:meanComparison}
    \end{figure}

The estimated mean surfaces for the two groups are plotted in Figure \ref{fig:meanComparison}.  The ASD group tends to have positive condition differentiation between trials 30 and 55, whereas the TD group tends to have positive condition differentiation in earlier trials. Positive condition differentiation is thought to be indicative of learning, so these results suggest that the TD group is learning at a faster rate than the ASD group.  However, even though qualitatively the surfaces look very different, there is a substantial amount of heterogeneity in the subject-level data, resulting in broad confidence bands around the mean, and perhaps suggesting that differential patterns of condition differentiation between ASD and TD groups are best explored considering both the mean and the covariance structure.

\begin{figure}[!p]
\begin{center}
\begin{tabular}{l}
ASD\\
\includegraphics[width=1\textwidth]{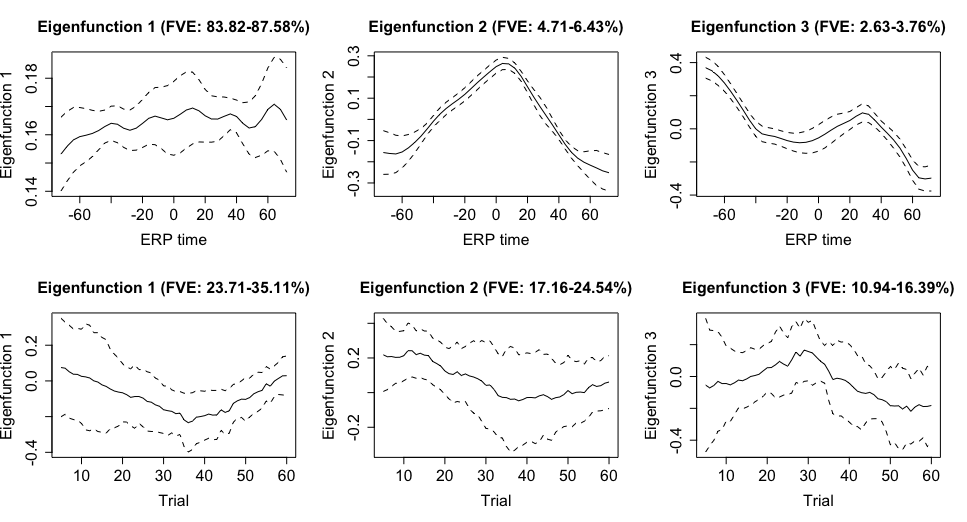} \\[.1in]
TD\\
\includegraphics[width=1\textwidth]{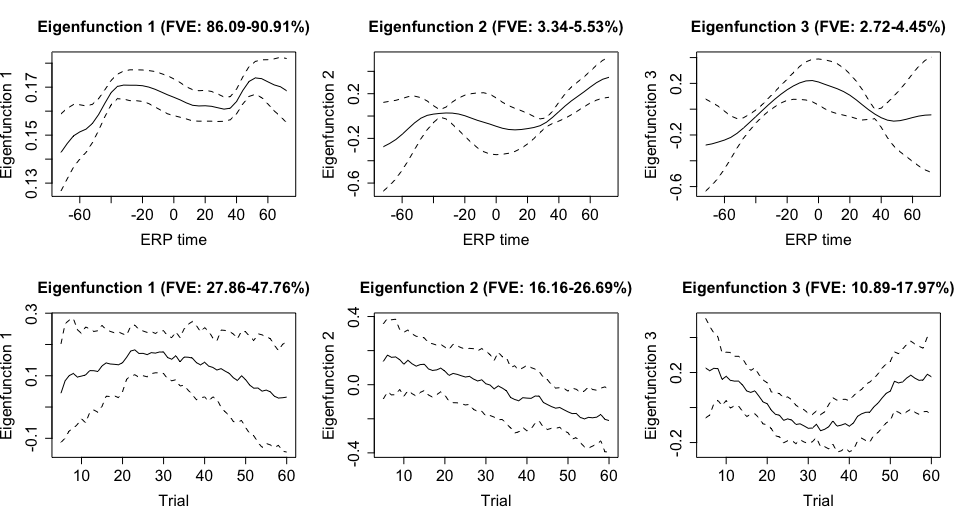}  \\
\end{tabular}
\end{center}
\caption{Marginal eigenfunctions with associated uncertainty for the ASD and TD groups. Solid black lines represent posterior means and dotted lines represent 95\% simultaneous credible bands.}
\label{fig:eigenASD}
\end{figure}

Next we conduct an eigen-analysis of the covariance structure for both cohorts separately. Figure \ref{fig:eigenASD} plots eigenfunctions of the marginal covariances over ERP time and trials. Credible intervals are calculated following \cite{Crainiceanu}.  

We start  by analyzing summaries indexing variability in ERP time. For both the TD and ASD cohorts, the first eigenfunction explains  the vast majority of the marginal covariance (84\%-88\% in ASD, and 86\%-90\% in TD).  In both groups this first eigenfunction is relatively flat and can be interpreted as representing variability in the overall level of condition differentiation within a trial. The magnitude and shape of variation is comparable between TD and ASD children. Finer differences may be detected in the second and third eigenfunction, which  further characterize variability in the shape of the ERP waveforms about the P300 peak. For both cohorts, however, these summaries represent only a small percentage of the variance in ERP waveform within trial.

Perhaps more interesting is an analysis of the marginal covariance across trial, as probabilistic learning patterns are likely to unfold with prolonged exposure to expected vs. unexpected shape pairings. For the ASD group, the first eigenfunction dips in an approximately quadratic fashion, suggesting  enhanced variability in condition differentiation at around trial 35. Similarly, for the TD group, the first trial eigenfunction has a slight peak around trial 25. A  possible interpretation of these covariance components relates to implicit learning, with higher variance in differentiation occurring earlier for TD than for ASD children. For both TD and ASD, the second eigenfunction across trials is can be interpreted as a contrast between high condition differentiation at early trials and low condition differentiation at later trials. Finally for the ASD cohort, the third eigenfunction exhibits a peak around trial 30. A possible interpretation would identify heterogeneity in the timing of learning, with some of the trajectories inducing variation in condition differentiation around trial 30, as opposed to the first eigenfunction identifying increased variance at around trial 35. Similarly for the TD group, the  third trial eigenfunction has a dip around trial 35, indexing delayed increased variability in  condition differentiation around trial 35.

\section{Discussion}
\label{s:discussion}
In this paper we provide a probabilistic characterization of longitudinal-functional data. As part of our work we propose a joint framework for the estimation of the mean or the regression function, and a flexible prior for covariance operators. Regularized estimation relies crucially on the projection of a set of basis coefficients onto a latent subspace, with adaptive shrinkage achieved via a broadly supported class of product Gamma priors. While we have not established theoretical results on posterior consistency, we have shown that the proposed framework exhibits competitive operating characteristics, when compared with alternative modeling strategies.

Importantly, uncertainty quantification, is achieved without having to rely on the asymptotic performance of bootstrap methods. From an applied perspective, analysts are charged with choosing the appropriate projection space. However, we see this as a feature rather than a problem, as different data scenarios may require and motivate the used of alternative basis systems. Because regularization is achieved jointly with estimation, inference is straightforward and does not need to account separately for the estimation of nuisance parameters or the choice of a finite number of eigenfunction to use in a truncated version of the model, as is the case for FPCA-based methods.  

We have shown that posterior inference using MCMC is implemented in a relatively straightforward fashion and need not rely on complicated posterior sampling strategies. When dealing with large data-sets, this na\"{i}ve inferential strategy may not be appropriate. For example,  the computation of marginal covariance functions $K_{\mathcal{S}}(s,s')$ and $K_{\mathcal{T}}(t,t')$ can be slow for designs with many longitudinal or functional time points, as numerical marginalization requires the computation of a four dimensional covariance function $K\{(s,t), (s',t')\}$. Some potential solutions include considering approximate  computation through  variational Bayes, or MAP approximation based on EM strategies.

From a modeling perspective, our probabilistic characterization of the longitudinal-functional covariance function is essentially equivalent to the weakly-separable model of  \cite{Chen2015}.  While more general than a strictly separable model, this strategy makes strong assumptions about the structure of a high-dimensional covariance operator. Testing strategies have been developed in the literature (\citealt{Chen2017}). However, we find that a more natural approach to the problem is one of regularized estimation. In this setting, a possible extension of our modeling framework could include an embedding strategy for the regularization of a non-separable covariance operator towards a weakly separable one.

\section{Software}
\label{sec5}

Software in the form of an R package including complete documentation and a sample data set is available from https://github.com/jshamsho/LFBayes

\section{Supplementary Material}
\label{sec6}

Supplementary material is available online at
\url{http://biostatistics.oxfordjournals.org}.

\section*{Acknowledgments}

{\it Conflict of Interest}: None declared.

\bibliographystyle{biorefs}
\bibliography{refs}

\newpage

\begin{table}
    \centering
    \begin{tabular}{ccccc}
    \hline
     Case 1 & & Bayes & Product & Empirical\\
   \hline
    \multirow{2}{*}{$n=30$} & $\mu(s,t)$ & .014 (.005, .038) &  .019 (.010, .044)& .019 (.010, .044)\\
    & $K\{(s,t),(s',t')\}$ & .062 (.023, .224) & .085 (.047, .200) & .151 (.097, .297)\\
    \multirow{2}{*}{$n=60$} & $\mu(s,t)$ & .007 (.003, .019) & .010 (.005, .021) & .010 (.005, .021)\\
    & $K\{(s,t),(s',t')\}$ & .030 (.010, .097) & .057 (.038, .128) & .076 (.050, .151)\\
    \hline
    \multicolumn{5}{l}{Case 2}\\
    \hline
    \multirow{2}{*}{$n=30$} & $\mu(s,t)$ & .024 (.007, .101) & .031 (.013, .118) & .031 (.013, .118)\\
    & $K\{(s,t),(s',t')\}$ & .039 (.011, .184) & .050 (.012, .202) & .067 (.030, .228)\\
    \multirow{2}{*}{$n=60$} & $\mu(s,t)$ & .014 (.004, .054) & .017 (.007, .062) & .017 (.007, .062)\\
    & $K\{(s,t),(s',t')\}$ & .019 (.005, .091) & .024 (.007, .093) & .032 (.014, .106)\\
    \hline
    \multicolumn{5}{l}{Case 3}\\
    \hline
    \multirow{2}{*}{$n=30$} & $\mu(s,t)$ & .155 (.046, .389) & .160 (.051, .393)  & .160 (.051, .393) \\
    & $K\{(s,t),(s',t')\}$ & .051 (.016, .187)& .051 (.014, .183) & .067 (.023, .200)\\
    \multirow{2}{*}{$n=60$} & $\mu(s,t)$ & .073 (.019, .216) & .076 (.021, .219)& .076 (.021, .219)\\
    & $K\{(s,t),(s',t')\}$ & .028 (.008, .091) & .027 (.007, .089) & .034 (.011, .099)\\
    \hline
    \end{tabular}
    \caption{Mean and covariance relative errors under under the three settings described in section \ref{s:simulation}. The two sample sizes used are $n = 30$ and $n = 60$.}
    \label{t:MeanCovSims}
\end{table}

\begin{table}
\centering
\noindent\begin{tabular}{lccc}
\hline
($p_1$, $p_2$) & DIC & BIC 1 & BIC 2\\
\hline
(5, 5) & 1.29 (1.20, 1.37) & 1.49 (1.40, 1.57)  & 1.59 (1.51, 1.67)\\
(10, 10) & 1.22 (1.14, 1.29) & 1.40 (1.32, 1.47) & 1.50 (1.43, 1.57)\\
(15, 15) & 1.24 (1.16, 1.31) & 1.41 (1.33, 1.48) & 1.52 (1.43, 1.59)\\
\hline
\end{tabular}
\caption{Information criteria for case 2. Each $(p_{1}, p_{2})$ combination is repeated 1000 times. The table reports the .5, .1, and .9 quantiles of the information criteria over 1000 simulation. Each number is on the $10^{4}$ scale.}
\label{t:case 2 info}
\end{table}

\newpage

\end{document}